\documentclass[3p,twocolumn]{elsarticle}
\usepackage{xspace}

\usepackage{lineno,hyperref}
\modulolinenumbers[5]

\newcommand{\Nmu}{N_\mu}

\newcommand{\Xmax}{X_{\rm max}}

\newcommand{\Elab}{E_{\rm Lab}}
\newcommand{\Esat}{E_{\rm sat}}
\newcommand{\Escale}{E_{\rm scale}}
\newcommand{\EGS}{E_{\rm GS}}
\newcommand{\Eth}{E_{\rm th}}
\newcommand{\sigmatot}{\sigma_{\rm tot}}
\newcommand{\sigmael}{\sigma_{\rm el}}
\newcommand{\sigmainel}{\sigma_{\rm inel}}

\journal{Physics Letters B}

\begin{document}

\begin{frontmatter}

\title{ Proton-proton interactions at the Highest Energies: \\
the Grey-Disk and the Core-Corona models
}

\author[LIP,IST]{R. Concei\c{c}\~ao}
\author[LIP,IST]{M. Pimenta}



\address[LIP]{LIP - Laborat\'orio de Instrumenta\c{c}\~ao e F\'isica Experimental de Part\'iculas, Lisbon, Portugal}
\address[IST]{Departamento de F\'isica, Instituto Superior T\'{e}cnico, Universidade de Lisboa, Lisbon, Portugal}

\begin{abstract}

In this letter, the Grey-Disk and the Core-Corona models are combined in a scenario that aims to explain different unexpected features observed in the interactions of the highest energy particles in the Earth's atmosphere. In particular, the observed distributions of  $\Xmax$ and $\Nmu$ are explained, assuming that the ultra-high-energy cosmic ray spectrum is dominated by protons produced by a few extra-galactic sources with different injection spectra and maximum energy cutoffs. 
The compliance of this heterodox scenario with other relevant observations as 
the instep region in the energy spectrum, the event-by-event correlations between $\Xmax$ and the detected signal at the ground and the absence of detection, so far, of UHE neutrinos or photons, are also briefly discussed.
If confirmed, such a scenario would dramatically change the present understanding of the UHECR primary composition and, thus, the characteristics of their sources.

\end{abstract}

\end{frontmatter}


\section{Introduction}

For many years it was commonly believed that at the highest energies (above $10^{19}\,$eV),  protons were probably the main component of UHECR being its arrival flux at Earth highly suppressed due to the existence of the  GZK mechanism \cite{GZK1,GZK2}. Such an idea was strongly questioned by the measurements of the distributions of the maximum development ($\Xmax$) of the  Extensive Air Showers (EAS)\cite{Xmax1,Xmax2}; by the distributions of the number of muons ($\Nmu$) detected at the ground~\cite{MuonsAverage,MuonsAMIGA,MuonsFluctuations}, and, more recently, by the discovery of an instep steepening region in the energy spectrum around $10^{19}\,$eV \cite{instep}.

 The net result is that, nowadays, it is believed that at the highest energies, UHECR composition is dominated by the superposition of increasing extra-galactic heavier nuclei whose energy spectra are sharply suppressed by the limited maximum energy that the sources may provide~\cite{AugerSpectrumFit}. Furthermore, the recent measurements of the event-by-event correlations between $\Xmax$ and the detected signal at the ground water Cherenkov detector stations \cite{XmaxS1000} excluded, in principle, with a high significance, a pure mass composition at these high energies.

In this letter, we argue that a scenario where the UHECR spectrum would be dominated by protons produced by a few extra-galactic sources with different injection spectra and maximum energy cutoffs is not excluded. 

The possibility that, at high energies, the measurements of the moments of the $\Xmax$ distributions ($\left< \Xmax \right>$, $\sigma({\Xmax})$) could be explained by a fast growth of the proton-proton total cross-section was introduced in the past~\cite{ppcross}. Namely, in a simple Grey-Disk model \cite{BlackDisk2}, the rapid increase of the cross-section was associated with a fast rise in the matter density inside the proton (the disk opacity). 
    
On the other hand, the state-of-the-art hadronic interaction models~\cite{EPOS1,EPOS2,sib,sib23d,QGS1,qgs2}, which have been tuned to the LHC data, cannot provide a consistent solution for the measured shower observables in terms of mass composition, displaying a muon deficit in the EAS simulations~\cite{PRLGlennys,PRLAugerMuons,Jakub}. To cope with this deficit, a new class of phenomenological models, designated by  Core-Corona model has recently been introduced  \cite{CoreCorona1,CoreCorona2,CoreCorona3}. In these models, two different particle production mechanisms may exist at the highest energies: the usual string fragmentation, corresponding to the more peripheral collisions (corona); and collective statistical hadronization, corresponding to the more central collisions (core), producing lower electromagnetic energy fractions and consequently higher numbers of muons.

Both the Grey-Disk and the Core-Corona models, which aim to explain two apparently different unexpected features of the interactions of the highest energy particles in the Earth atmosphere, are briefly described in the section~\ref{sec:BD-CoreCorona}, where their possible connections are also discussed.   

In a scenario where the changes on the shower observables are caused mainly by changes in the hadronic interaction properties, the present understanding of the UHECR primary composition would change drastically and, thus, the characteristics of their astrophysical sources. Furthermore, it might also give a simple explanation of the difference between the Auger and the Telescope Array (TA) data at the far end of their energy spectra \cite{AugerTA}, putting the focus on the variability of the sources. These aspects are briefly developed in section~\ref{sec:discussionandconclusions}.

%

\section{The Grey-Disk and the Core-Corona models}
\label{sec:BD-CoreCorona}

In the Grey-Disk model~\cite{BlackDisk1,BlackDisk2} the elastic scattering amplitude at a given square of the centre-of-mass energy, $s$, is defined, in the impact parameter space, $b$, as equal to a positive real constant $\overline{\Omega}$ (the disk opacity),  for any $b$ value lower than the disk radius $R$, and equal to zero otherwise. 
Both $\overline{\Omega}$ and $R$ are growing functions of $s$.

In this model:

\begin{equation}
\label{eq:greydisktot}
\sigmatot (s)=  2 \pi (1 - e^{- \overline{\Omega}(s)}) R^2(s), 
\end{equation}

\begin{equation}
\label{eq:greydiskcel}
\sigmael (s)=   \pi (1 - e^{- \overline{\Omega}(s)})^2 R^2(s), 
\end{equation}

\begin{equation}
\label{eq:greydiskinel}
\sigmainel(s)=   \pi (1 - e^{- 2 \overline{\Omega}(s)}) R^2(s),
\end{equation}

\begin{equation}
\frac{\sigmael(s)}{\sigmatot(s)} = \frac{1}{2} \left( 1 - e^{- \overline{\Omega} (s)}\right) = \frac{1}{2} \overline{\Gamma} (s).
\label{eq:sigmael_tot}
\end{equation}

where $\sigmatot$, $\sigmael$ and $\sigmainel$ are, respectively, the cross-sections for the total, elastic and inelastic processes,  and $\overline{\Gamma}(s)$ is designated hereafter as the mean profile of the disk.

Therefore, $\sigmatot(s)$ and $\sigmael(s)$ can rise either by increasing $R(s)$ or by increasing $\overline{\Omega}(s)$, while the ratio $\sigmael(s)/\sigmatot$ is just a function of $\overline{\Omega}(s)$. 

In the limit $\overline{\Omega} \rightarrow \infty$, known as the black disk limit: 
$\overline{\Gamma}(s)=(1 - e^{- \overline{\Omega}(s)})\rightarrow 1$ ;
$\sigmatot (s)=  2 \pi  R^2(s)$;  $\sigmainel(s)=   \pi R^2(s)$; 
$\frac{\sigmael(s)}{\sigmatot(s)} = \frac{1}{2} $.

In reference \cite{BlackDisk2}, all the available proton-proton cross-sections measurements at the moment of the publication, from ISR ($\sqrt{s}\sim 20-60\,$GeV) to LHC ($\sqrt{s} \sim 7\,$TeV),
were found to be well described by a Grey-Disk model where:

\begin{itemize}
\item 
the radius was increasing monotonically: 

\begin{equation}
R(s) = R_0 + \beta \log \left( \frac{s}{s_0} \right).
\label{eq:radius_term}
\end{equation}
with, $R_0 = 2.52 \pm 0.02\,$GeV$^{-1}$;  $\beta=0.067\pm0.002\,$ GeV$^{-1}$; $\sqrt{s_0} = 10\,$GeV,

\item 
 the opacity had two different regimes:
\begin{itemize}
\item 
constant up to $\sqrt{s}\sim 100\,$GeV (corresponding to the Geometric Scaling (GS) regime \cite{ref8_1,ref8_2}
), 

\begin{equation}
\overline{\Omega} = \textrm{const} = 0.4298 \pm 0.0024 ;
\label{eq:Omegaconst}
\end{equation}

\item 
increasing with the beam rapidity  ($Y=\log \left( \sqrt{s} / m_p \right)$)
for higher $\sqrt{s} $,

\begin{equation}
\overline{\Omega}(s) = \frac{2}{k} \lambda Y,
\label{eq:density}
\end{equation}

where,   $k = 5.52 \pm 0.15$ and $\lambda = 0.23 \pm 0.009$. 
\end{itemize}

\end{itemize}

While in the last years the proton-proton cross-section has been measured at the Large Hadron Collider up to energies of $\sqrt{s} = 13\,$TeV~\cite{TOTEM_Xsec}, the model proposed in~\cite{BlackDisk2} is still able to describe this data without re-tuning its parameters.

In the same reference, \cite{BlackDisk2}, to describe the unexpected behaviour of the $\Xmax$ moments (mean and RMS) observed at Auger, it was proposed a sharp
\footnote{While the breaks indicate different regimes, in the same reference ~\cite{BlackDisk2}, 
the authors were able to reproduce all the collect cross-section data with a regular $C^1$ function by evolving the $k$ parameter with the beam rapidity.} 
increase in the opacity reaching the Black Disk limit at $\sqrt{s}\sim 100\,$TeV. 

\medskip

In the Core-Corona model \cite{CoreCorona1}, above a certain centre-of-mass energy, a quark-gluon plasma (QGP) may be formed in the central region -- \emph{ the core} -- of proton-proton collisions and its statistical hadronization is responsible by a fraction ($w_{\rm core}$) of the produced outgoing particles. The remaining fraction of the outgoing particles ($1 - w_{\rm core}$) are produced, in this model, by string fragmentation in the more peripheral region --\emph{the corona}. In this way, high energy proton-proton collisions would mimic, somehow, heavier particle collisions where such behaviour was first predicted and observed.  

Taking $N_i$ as the total number of produced particles of the type $i$ and $N_i^{\rm core}$ and $N_i^{corona}$ the corresponding numbers of particles produced in the core and the corona region, respectively, one can write:

\begin{equation}
N_i = w_{\rm core} N_i^{\rm core} + (1 - w_{\rm core}) N_i^{\rm corona}.
\end{equation}

Note that both $N_i^{\rm core}$ and $N_i^{corona}$ are the multiplicities assuming that the hadrons involved in the collision are only made of core or corona, respectively. In these conditions, the quantity $w_{\rm core}$ is, therefore, the key parameter of the model and was set, in a fixed-target proton-proton collision, to:

\begin{equation}
 w_{\rm core} =  f_{w} \frac{\log_{10}(\Elab/\Eth)} {\log_{10}(\Escale/\Eth) }
 \label{eq:wcore}
\end{equation}

where, $\Elab$ is the beam energy and $\Eth$ and $\Escale$ are the extreme beam energies 
corresponding, respectively, to  $w_{\rm core}=0$ and  $w_{\rm core}=f_{w}$.  
$\Elab$ should also be $\ge \Eth $ and, whenever $\Elab \ge \Escale$,  $w_{\rm core}$ was set to $f_{w}$.

\begin{figure}[htp]
    \centering
   \includegraphics[width = 0.5\textwidth]{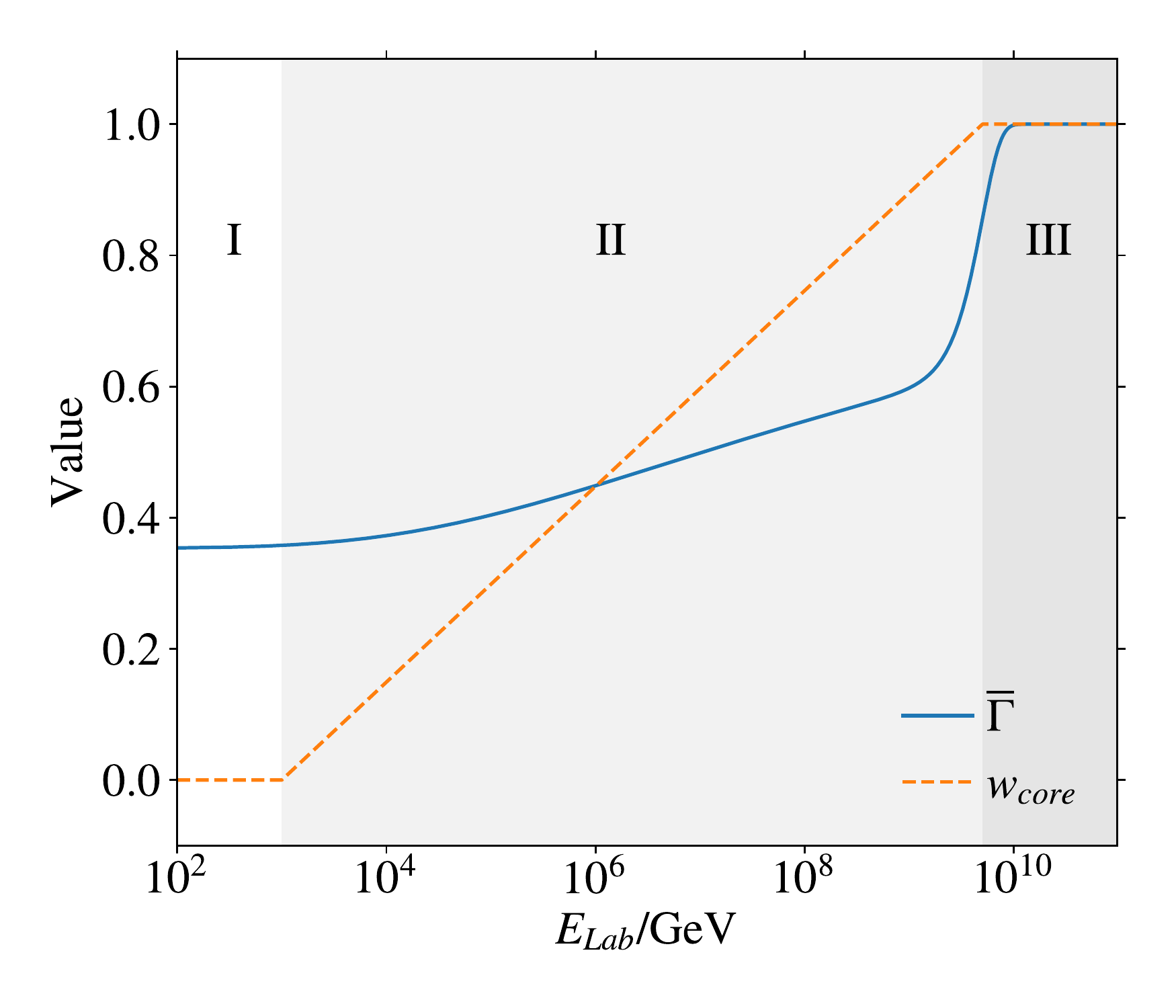}
    \caption{Energy evolution of the variables $\overline{\Gamma}$ (Grey-Disk Model) and  $w_{\rm core}$ (Core-corona Model), see text for details.}
    \label{fig:densitypar}
\end{figure}

\medskip

In both models, the Grey-Disk and the Core-Corona, the energy/matter density has the central role.
In the Grey-Disk the degree of absorption of the incoming wave is controlled by  $\overline{\Gamma} (\overline{\Omega}(s))$ (see equation \ref{eq:sigmael_tot}), which should be a function of the matter density. In the Core-Corona model $w_{\rm core}$ represents somehow the fraction of the collision area where the formation of QGP has occurred, and thus where the energy density is above some limit. 

Moreover, in these models, there are two extreme energies defining the three relevant phenomenological regions: 
\begin{itemize}
\item Region I - At the lowest energies, in the Core-Corona model, there is no formation of QGP ($\Elab \le \Eth$, $w_{\rm core} = 0$), and in Grey-Disk model the opacity is constant ( $ \Elab \le \EGS$, the GS regime); 
\item Region II - At intermediate energies there is a $\log \left( \sqrt{s} \right)$ excursion of the corresponding relevant parameters, $w_{\rm core}$ and $\overline{\Omega}$;
\item Region III - At the highest energies in the Core-Corona model, QGP is always formed ($ \Elab \ge \Escale$, $w_{\rm core} = 1$), and in Grey-Disk model the opacity $\overline{\Omega}$ is infinity ($ \Elab \ge \Esat$, the Black Disk).
\end{itemize}

Thus, it is phenomenologically tempting to connect both models establishing a relation between these two key variables, $\overline{\Gamma} (\overline{\Omega}(s))$ and $w_{\rm core}(s)$, connecting in this way the variation of the cross-section, above the  GS regime, with the increase of  the  number of produced muons in the hadronization  phase 
due to the formation of QGP. 

A more solid theoretical approach should be able to establish the function $w_{\rm core} = f(\overline{\Gamma})$.  Nevertheless, 
one may postulate the equivalence of the two extreme energies in both models, such that $\Eth \sim \EGS \sim 10^{3}\,$GeV; $\Escale \sim \Esat \sim 5 \times 10^{9}\,$GeV, and  verify that, setting $f_{w}=1$ in equation \ref{eq:wcore}, the two variables have  not only an equivalent  behaviour as a function of $\sqrt{s}$ but also that their numerical values are similar  (see figure \ref{fig:densitypar}). Furthermore, these values are
well inside the phase space of the parameters of the Core-Corona model, for which there is a significant impact on the increase of the mean number of muons in proton EAS showers (see reference \cite{CoreCorona1}).

\section{Discussion and Conclusions}
\label{sec:discussionandconclusions}

The conjunction of the two models may reproduce, as discussed in the previous section, the main features  observed in the distributions of $\Xmax$ and $\Nmu$ at the highest energies. Other usually quoted distributions that may indicate an a priori unexpected composition scenario are the energy spectrum; the event-by-event correlations between $\Xmax$ and the detected signal at the ground; the non-observation, so far, of ultra-high energy neutrinos or photons; the degree of observed anisotropies of the cosmic rays arrival directions at the highest energies.

The instep region observed recently in the energy spectrum between $1.3 \times 10^{19}\,$eV and  $5 \times 10^{19}\,$eV~\cite{instep},  has been interpreted using models where the acceleration of the particles in the sources depends only on their rigidity, which would be translated as an energy-dependent mass composition. Although very appealing, these models imply an unexpected, very hard energy spectrum at the sources so that each mass component will be dominant in successive small energy intervals. In fact, if there would be a smooth transition between the several mass components, the fluctuations of the $\Xmax$ distributions would increase, in contradiction to its monotonic decrease as a function of energy above  $5 \times 10^{18}\,$eV.

An alternative interpretation, compatible with the scenario of proton-dominant sources, would be that the observed instep region would result from the superposition of the spectra of a small number of extra-galactic sources, each with a slightly different maximum energy cutoff and/or spectral indices. This interpretation would also explain the fact that, at high energies, the energy spectra measured by Auger and TA collaborations have, within the systematic errors of the energy scales, a good agreement for arrival directions in the common declination band ( $-15^\circ < \delta < 24.8^\circ $) with a steepening of the spectrum at around $5 \times 10^{19}\,$eV, while, for higher declinations, only accessible to TA, the observed steepening is around $7 \times 10^{19}\,$eV.

The no observation, so far, of UHE neutrinos or photons disfavours, although not yet excludes, the initial GZK proton scenario. However, one should note that in the alternative scenario suggested in the present letter, the steepening of the proton spectrum is due to the limited maximum energy that the accelerating sources may provide and not by the GZK mechanism acting on protons produced in far-away sources.

\begin{figure}[htp]
    \centering
    \includegraphics[width = 0.9\columnwidth]{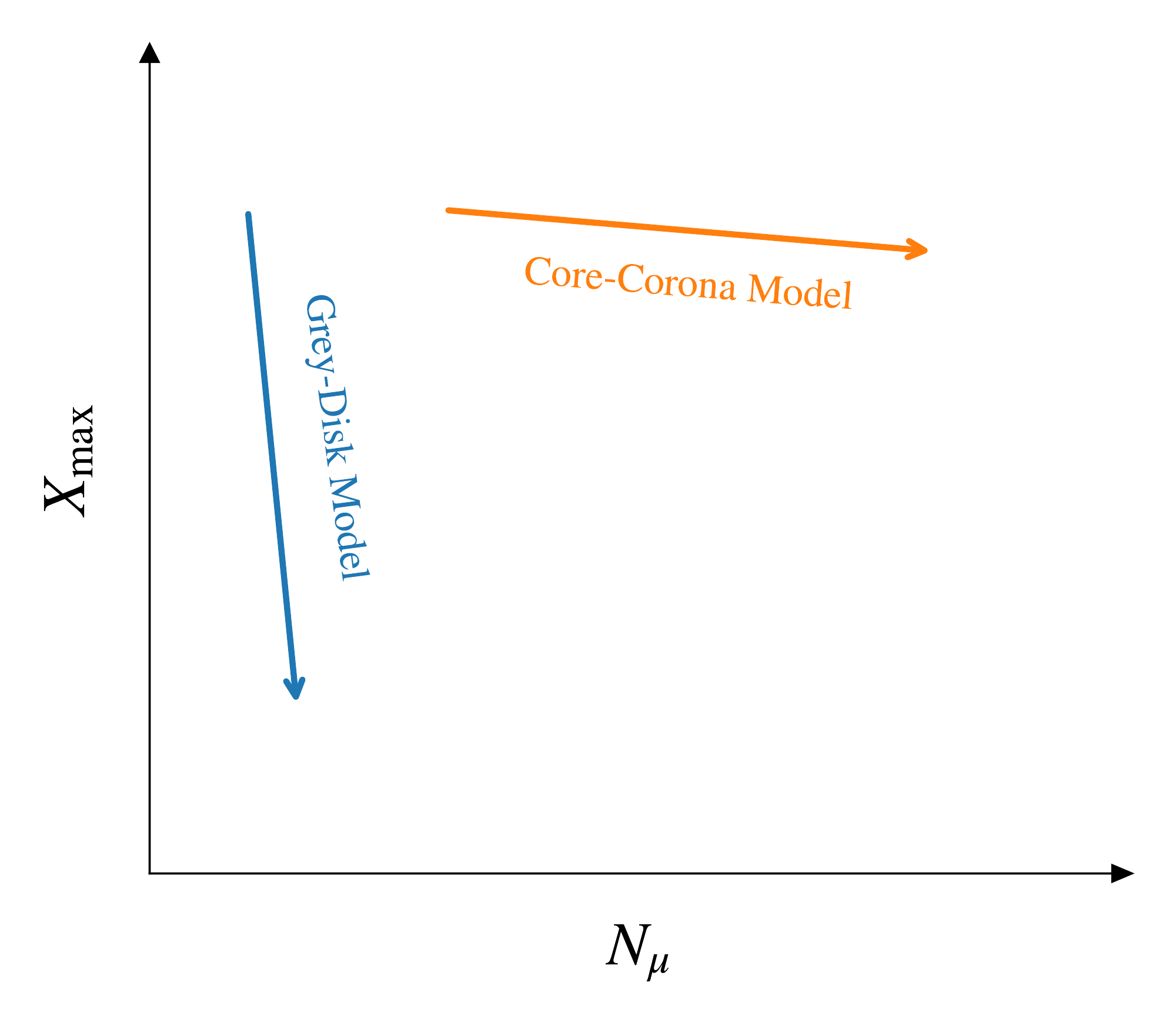}
    \caption{Scheme of the plane $(\Xmax , \Nmu)$ and the changes caused by the Grey-Disk and Core-Corona model evolution with the primary energy. The size of the arrows is arbitrary.}
    \label{fig:XmaxNmu}
\end{figure}

The event-by-event correlations between $\Xmax$ and the detected signal at the ground ($S_{38}^*$) have been claimed as evidence that in the energy bin between  $10^{18.5}\,$eV and $10^{19}\,$eV any pure mass composition would be excluded. However, this is precisely the energy region where, in the Grey-Disk model, there is a sharp increase of the opacity which is translated in the rise of the total cross-section (shallower $\left< \Xmax \right>$) and, in the Core-Corona model there is a significant increase of the number of muons (higher $S_{38}^*$). In these terms,  the correlation between the two variables should thus be negative, as observed. It should also be noted that these models influence these shower observables differently (see figure~\ref{fig:XmaxNmu}). On the one hand, changes on the hadron-air cross-section significantly impact the mean value of $\Xmax$ while the number of muons at the ground remains essentially untouched. On the other hand, Core-Corona models have been shown to be able to increase the EAS muon content with little effect on the $\Xmax$.  As such, the correlation factor for these models can significantly differ from the expectations for mass compositions under the use of standard hadronic interaction models.

Finally, a disclaimer should be made. Our purpose in this letter was not to prove the superiority of a heterodox scenario combining the Grey-Disk and the Core-Corona models but to widen the discussion that had been held in the last tens of years without, so far, reaching, in our opinion,  complete satisfactory answers. Extensive simulations, which are out of the scope of this letter, need to be made to cope with all the aspects enumerated above. Additionally, such studies should consider the results of the anisotropy studies on the UHECR arrival directions, which are currently favouring a limited number of high energy sources~\cite{instep}. The studies contain many assumptions, for instance, on the extra-magnetic field strength and coherence. As such, only through a combined study will we be able to build a consistent picture that helps us to understand the nature and origin of UHECR.

\section*{Acknowledgements}
We would like to thank Alan Watson, Antonio Bueno, Jaime Alvarez-Mu\~niz, Jakub Vicha, Roger Clay, Sofia Andringa and Tanguy Pierog for carefully reading the manuscript and for useful comments.
The authors also thank for the financial support by OE - Portugal, FCT, I. P., under project CERN/FIS-PAR/0020/2021. R.~C.\ is grateful for the financial support by OE - Portugal, FCT, I. P., under DL57/2016/cP1330/cT0002. 

\paragraph{Dedication} We dedicate this article to the memory of our friend, Prof. Jorge Dias de Deus, who passed away in February 2021, and with whom one of us (M. Pimenta) has worked closely during many years, over the last two decades of the last century to the end of the first decade of this century, developing new phenomenological ideas in the field of high energy hadronic interactions and, namely, in the Black Disk interpretation of the first Auger data.

\bibliography{references}

\end{document}